\documentclass[aps,preprint,pra,aps,amssymb]{revtex4}
\usepackage[]{graphicx}
\usepackage{subfigure}

\begin{document}

\title{Heating of the fuel mixture due to viscous stress ahead of accelerating flames
in deflagration-to-detonation transition}

\author{Damir Valiev$^{1,2}$, Vitaly Bychkov$^{1}$,
V'yacheslav Akkerman$^{1}$, Lars-Erik Eriksson$^{3}$\\ and Mattias
Marklund$^{1}$}

\affiliation{$^{1}$Department of Physics, Ume\aa\ University,
SE--901 87, Ume\aa, Sweden
\\
$^{2}$Department of Materials Science and Engineering, Royal
Institute of Technology, SE--100 44 Stockholm, Sweden
\\
%$^{3}$Institute for High Energy Densities JIHT RAS, Izhorskaya St.
%13/19, Moscow 125 412, Russia
%$^{4}$Combustion Technology, Department of Mechanical Engineering,
%Eindhoven Technical University P.O.Box 513 5600 MB Eindhoven, the
%Netherlands
$^{3}$Department of Thermo- and Fluid Dynamics, Chalmers
University of Technology, SE--412 96 G\"{o}teborg, Sweden}

\begin{abstract}
The role of viscous stress in heating of the fuel mixture in
deflagration-to-detonation transition in tubes is studied both
analytically and numerically. The analytical theory is developed
in the limit of low Mach number; it determines temperature
distribution ahead of an accelerating flame with maximum achieved
at the walls. The heating effects of viscous stress and the
compression wave become comparable at sufficiently high values of
the Mach number. In the case of relatively large Mach number,
viscous heating is investigated by direct numerical simulations.
The simulations were performed on the basis of compressible
Navier-Stokes gas-dynamic equations taking into account chemical
kinetics. In agreement with the theory, viscous stress makes
heating and explosion of the fuel mixture preferential at the
walls. The explosion develops in an essentially multi-dimensional
way, with fast spontaneous reaction spreading along the walls and
pushing inclined shocks. Eventually, the combination of explosive
reaction and shocks evolves into detonation.
\end{abstract}

%\pacs{}

\maketitle

%========================================================================%
%                           1. Introduction                              %
%========================================================================%
\section{Introduction}
Deflagration-to-detonation transition (DDT) is one of the most
important and least understood problems in combustion science. It
is well known that a deflagration wave propagating from a closed
tube end may spontaneously accelerate and trigger an explosion in
the fresh fuel mixture, which goes over to detonation
\cite{Shelkin-1940,Williams-1985,Zeldovich.et.al-1985,Shepherd.et.al-1992,
Roy.et.al-2004,Oran&Gamezo-2007}. The mechanism of DDT was
described qualitatively already by Shelkin \cite{Shelkin-1940}.
According to the Shelkin explanation, thermal expansion in the
burning process produces a strong flow in the unburnt fuel
mixture, which becomes non-uniform because of non-slip at the tube
walls. The non-uniform flow makes the flame curved, which
increases the flame surface and hence the volumetric burning rate;
it makes the flow even stronger and leads to flame acceleration.
An accelerating flame pushes compression waves and shocks, the
temperature in the fuel mixture increases, the reaction goes even
faster, until explosion happens ahead of the flame front and
evolves into detonation.

For a long time it was a general belief that DDT is impossible
without artificial turbulence generation. Since turbulence and
turbulent burning are not so well-understood yet, it presents a
major obstacle in developing a theory of flame acceleration and
DDT. Only recently a constructive idea was suggested that DDT may
be achieved even in a laminar flow in tubes with adiabatic walls
\cite{Kagan&Sivashinsky-2003,Ott.et.al-2003}. Though it is
difficult to obtain adiabatic walls in a real experiment, these
conditions may be easily imitated in direct numerical simulations
\cite{Kagan&Sivashinsky-2003,Ott.et.al-2003}. Starting with this
idea, the analytical theory of laminar flame acceleration was
developed in \cite{Bychkov.et.al-2005,Akkerman.et.al-2005}; the
theory was quantified by extensive numerical simulations. The idea
of laminar flame acceleration allowed also direct numerical
simulations of DDT, e.g. see
\cite{Kagan&Sivashinsky-2003,Kagan.et.al-2006,Liberman.et.al-2006}
 and a recent review \cite{Oran&Gamezo-2007}. The simulations demonstrated
many interesting details of the DDT; at present it is not quite
clear, which of them are intrinsic to DDT, making them the
backbone of the process, and which are just supplementary effects
of minor importance. One of these details is heating of the fuel
mixture due to viscous stress, which makes the explosion
preferential close to the walls. This is different from the
classical Shelkin scenario of DDT with heating produced mainly by
the compression/shock waves.

A quantitative analytical theory of explosion triggering by an
isentropic compression wave ahead of an accelerating flame was
developed recently in \cite{Bychkov-Akkerman-2006}. This theory
disregarded influence of the viscous stress. However, an important
role for viscous stress is in line (at least, ideologically) with
the model of the DDT, based on hydraulic resistance, see
\cite{Brailovsky-Sivashinsky-2000} and references therein. This
model had another shortcoming; it was one-dimensional and did not
take into account Shelkin acceleration, which is an intrinsic part
of DDT. Thus we came to the question, how strong the role of
viscous stress is in heating of the fuel mixture and in DDT,
taking into account the multi-dimensional Shelkin mechanism of
this process. The purpose of the present paper is to answer this
question. The paper considered only the geometry of smooth tubes,
with no obstacles at the walls.

We studied the role of viscous stress both analytically and
numerically. The analytical study is based on the theory of
laminar flame acceleration \cite{Bychkov.et.al-2005}. The theory
works excellently for low Mach numbers, typical for the beginning
of flame acceleration, about  $Ma=10^{-3}-10^{-2}$. We found
temperature profile in the flow ahead of the flame front with
maximal temperature increase achieved at the walls. We show that,
at low values of the Mach number, the role of viscous heating is
extremely small, it scales as  $Ma^{2}$. By comparison, heating
due to the compression wave scales as  $Ma$. However, temperature
at the walls grows faster in time than at the axis. According to
the theoretical estimates, the heating effects of viscous stress
and of the compression wave become comparable as soon as the Mach
number (defined using total burning rate) reaches values above
0.6. Another interesting point is that heating due to viscous
stress does not depend on the Reynolds number, at least as long as
the Mach number is sufficiently low. When the Mach number becomes
about 0.05 and higher, the theory \cite{Bychkov.et.al-2005} works
only qualitatively, providing order-of-magnitude estimates.

In that range of Mach numbers we investigated the role of viscous
heating by direct numerical simulations. The simulations were
performed on the basis of compressible Navier-Stokes gas-dynamic
equations taking into account chemical kinetics. It was again
established that the fuel mixture is heated stronger at the walls
than along the tube axis. In agreement with the analytical theory,
we found that heating effects of viscous stress and compression
wave become comparable at large values of the Mach number; viscous
heating makes explosion of the fuel mixture preferential at the
walls. The explosion develops in an essentially multi-dimensional
way, with fast spontaneous reaction spreading along the walls and
pushing rather strong shocks inclined with respect to the tube
axis. Eventually, the combination of explosive reaction and shocks
evolves into detonation.

The present paper is organized as follows: in Section II we
develop the analytical theory of heating due to viscous stress; in
Section III we describe the details of the numerical simulations;
in Section IV we present and discuss the results obtained; the
paper is concluded with a short summary.

%========================================================================%
%                  II. Theory of viscous heating                     %
%========================================================================%
\section{Theory of heating due to viscous stress ahead of accelerating flames}

\begin{figure}
\includegraphics[width=\textwidth\centering]{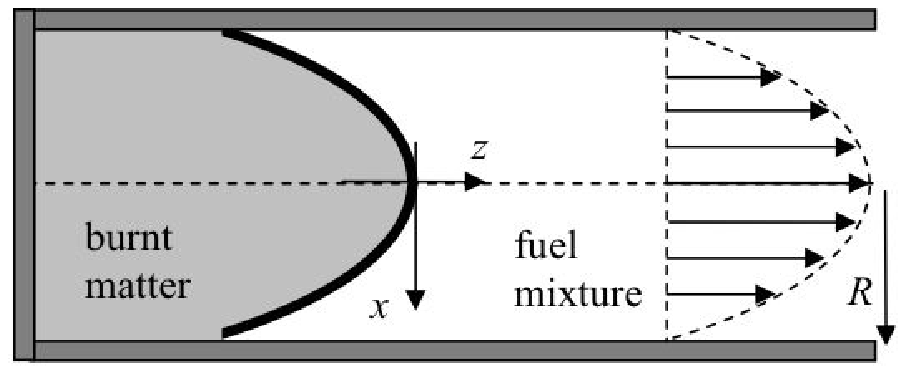}
 \caption{Flame propagating in a tube with non-slip walls.}
  \label{fig-1}
\end{figure}

We consider the same geometry as in \cite{Bychkov.et.al-2005}: a
laminar flame propagates in a two-dimensional (2-D) tube of
half-width  $R$ with adiabatic walls and non-slip at the walls as
shown schematically in Fig. \ref{fig-1}. Burning matter expands as
it passes the flame front; density ratio of the fuel mixture
$\rho_{f}$ and the burnt gas $\rho_{b}$  is typically rather
large,  $\Theta = \rho_{f}/ \rho_{b}= 5-8$. Because of the thermal
expansion, a flame propagating from the closed tube end pushes the
unburnt fuel mixture and generates a flow. It was shown in
\cite{Bychkov.et.al-2005} that the stream ahead of the flame may
be closely approximated by a plane-parallel flow along the walls
$\textbf{u} = \hat{\textbf{e}}_{z} u_{z}(x,t)$.

\begin{figure}
\includegraphics[width=\textwidth\centering]{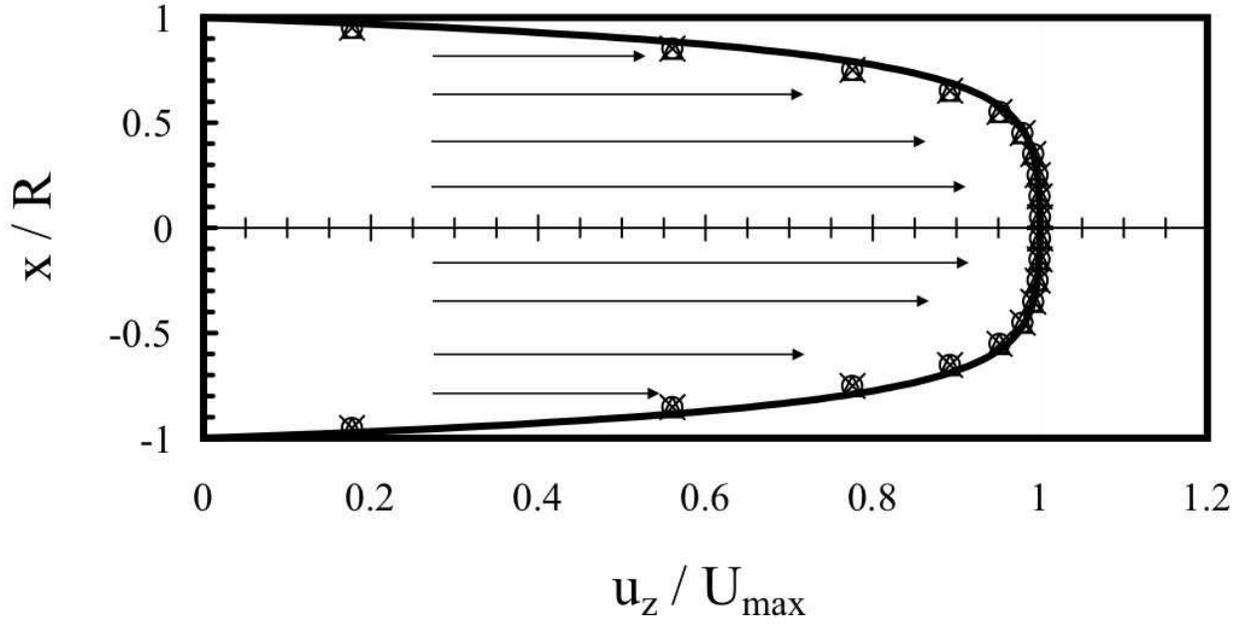}
 \caption{The velocity profile  $u_{z}$ scaled by the amplitude $U_{max}$.
The solid line shows the theoretical result Eq. (\ref{eq5}). The
markers correspond to the simulation results of
\cite{Bychkov.et.al-2005} for $Re = 50$,  $Pr = 0.5$ at the
distances $10R; 20R; 35R$ (circles, triangles and crosses) from the
flame at the time instance  $U_{f}t/R = 1.15$. The arrows illustrate
the direction of the flow.}
  \label{fig-2}
\end{figure}

The normal planar flame velocity $U_{f}$  provides a natural
scaling of velocity dimension for the problem. However, the total
burning rate $U_{w}$ is different from  $U_{f}$: it shows how much
fuel mixture is consumed per unit time by the whole flame front
and how much energy is produced. In the standard 2-D model of an
infinitely thin flame front, the relative increase in the burning
rate is simply equal to the increase in the total length of the
flame front. The propagating flame generates a flow, which is not
uniform because of non-slip at the walls. Friction retards the gas
close to the walls, while flow velocity at the tube axis is larger
than the average one. The non-uniform velocity profile distorts
the flame shape, which leads to increase in $U_{w}$ and hence to
flame acceleration. At sufficiently small values of the Mach
number, neglecting the starting transitional period, the flame has
been shown to accelerate exponentially \cite{Bychkov.et.al-2005}
%=========================== Equation (2) ==============================%
\begin{equation}
U_{w}/U_{f}=\exp\left(\sigma{U}_{f}t/R\right).
 \label{eq2}
\end{equation}
%=======================================================================%
For 2-D flow, the dimensionless acceleration rate $\sigma$  was
found to be \cite{Bychkov.et.al-2005}
%=========================== Equation (3) ==============================%
\begin{equation}
\sigma=\frac{(Re-1)^{2}}{4Re}\left(\sqrt{1+\frac{4Re\Theta}{(Re-1)^{2}}}-1\right)^{2},
 \label{eq3}
\end{equation}
%=======================================================================%
where the value  $Re = U_{f}R/\nu$ plays the role of the Reynolds
number for the problem. For large values of the Reynolds number,
Eq. (\ref{eq3}) predicts decrease of the acceleration rate
%=========================== Equation (4) ==============================%
\begin{equation}
\sigma=\Theta^{2}/Re.
 \label{eq4}
\end{equation}
%=======================================================================%
In \cite{Bychkov.et.al-2005}, the Navier-Stokes equation was
solved as
%=========================== Equation (5) ==============================%
\begin{equation}
u_{z}/U_{f}=(\Theta-1)\exp\left(\sigma{U}_{f}t/R\right)
\frac{\cosh \mu - \cosh (\mu x/R)}{\cosh \mu - \mu^{-1} \sinh
\mu},
 \label{eq5}
\end{equation}
%=======================================================================%
where  $\mu = \sqrt{\sigma Re}$. For large values of the Reynolds
number we have $\mu = \Theta$, and the flow resembles
qualitatively a combination of two boundary layers at the walls
separated by the main almost uniform stream, see Fig. \ref{fig-2}.
Paper \cite{Bychkov.et.al-2005} demonstrated a very good agreement
of the theory with direct numerical simulations of acceleration
flames.

Here we find the temperature increase in the fuel mixture ahead of
the accelerating flame using the results above. Similar to
\cite{Bychkov.et.al-2005}, we adopt the limit of incompressible
flow, which holds in the case of a sufficiently small Mach number.
In that case the equation of thermal conduction in the fuel
mixture is \cite{Landau&Lifshitz-1989}
%=========================== Equation (6) ==============================%
\begin{equation}
\frac{\partial T}{\partial t} + \textbf{u}\cdot \nabla T = \chi
\nabla^{2}T + \frac{\nu}{2C_{P}}\left(\frac{\partial
u_{i}}{\partial x_{k} } + \frac{\partial u_{k}}{\partial x_{i}
}\right)^{2} .
 \label{eq6}
\end{equation}
%=======================================================================%
In the case of a shear flow  $\textbf{u} = \hat{\textbf{e}}_{z}
u_{z}(x,t)$, with  $T=T(x,t)$, Eq. (\ref{eq6}) reduces to
%=========================== Equation (7) ==============================%
\begin{equation}
\frac{\partial T}{\partial t} = \chi \frac{\partial^{2}T}{\partial
x^{2}} + \frac{\nu}{C_{P}}\left(\frac{\partial u_{z}}{\partial x }
\right)^{2} .
 \label{eq7}
\end{equation}
%=======================================================================%
We introduce standard scalings  $R$, $U_{f}$  and  $R/U_{f}$ as
units of length, velocity and time. In that case we work with the
dimensionless values  $(\eta; \xi) = (x; z)/R$,  $\tau =
U_{f}t/R$, $w = u_{z}/U_{f}$, and the dimensionless temperature
$\vartheta = T/T_{0} - 1$. Taking into account the relation for
sound speed in a politropic gas $c_{s0}^{2}= (\gamma - 1)
C_{P}T_{0}$ with heat capacity $C_{P}$ and adiabatic exponent
$\gamma$, we rewrite Eq. (\ref{eq7}) as
%=========================== Equation (8) ==============================%
\begin{equation}
\frac{\partial \vartheta}{\partial \tau} = \frac{1}{PrRe}
\frac{\partial^{2}\vartheta}{\partial \eta^{2}} + (\gamma -
1)\frac{Ma_{0}^{2}}{Re}\left(\frac{\partial w}{\partial \eta }
\right)^{2} .
 \label{eq8}
\end{equation}
%=======================================================================%
where $Ma_{0}=U_{f}/c_{s0}$  is the Mach number defined with the
help of the planar flame velocity. The last term in Eq.
(\ref{eq8}) is specified by the theory \cite{Bychkov.et.al-2005}
with
%=========================== Equation (9) ==============================%
\begin{equation}
\left(\frac{\partial w}{\partial \eta } \right)^{2} = (\Theta -
1)^{2}\mu^{2} \exp (2\sigma \tau) \frac{\sinh^{2}
(\mu\eta)}{(\cosh\mu - \mu^{-1}\sinh \mu)^{2}}.
 \label{eq9}
\end{equation}
%=======================================================================%
We evaluate the possible effect of heating keeping in mind the
limit of large Reynolds number with  $\sigma = \Theta^{2}/Re$,
$\mu = \Theta$, see Eq. (\ref{eq4}). Maximal heating happens close
to the wall. Due to symmetry, we may consider only the wall at
$\eta = 1$, where $\sinh(\mu\eta) \approx \cosh (\mu\eta) \approx
\frac{1}{2}\exp(\mu\eta)$. Such an approximation holds with
exponential accuracy $\exp(-\Theta)<<1$. In that case we have
close to the wall (at $1-\eta << 1 $)
%=========================== Equation (10) ==============================%
\begin{equation}
\left(\frac{\partial w}{\partial \eta } \right)^{2} =
\frac{(\Theta - 1)^{2}}{(\mu - 1)^{2}}\mu^{4} \exp \left[2\sigma
\tau + 2\mu(\eta-1)\right] ,
 \label{eq10}
\end{equation}
%=======================================================================%
and $\partial w / \partial\eta \approx 0$ at the channel axis
$\eta = 0$. Then Eq. (\ref{eq8}) reduces to
%=========================== Equation (11) ==============================%
\begin{equation}
\frac{\partial \vartheta}{\partial \tau} = \frac{1}{PrRe}
\frac{\partial^{2}\vartheta}{\partial \eta^{2}} + (\gamma -
1)\frac{Ma_{0}^{2}(\Theta - 1)^{2}}{Re(\mu - 1)^{2}}\mu^{4} \exp
\left[2\sigma \tau + 2\mu(\eta-1)\right].
 \label{eq11}
\end{equation}
%=======================================================================%
A particular solution to Eq. (\ref{eq11}) may be written as
$\vartheta_{p} \propto \exp \left[2\sigma \tau +
2\mu(\eta-1)\right]$, which leads to
%=========================== Equation (12) ==============================%
\begin{equation}
\vartheta_{p} = - \frac{\left(\gamma-1\right)Pr\left(\Theta -
1\right)^{2}\mu^{2}}{2\left(2-Pr\right)\left(\mu - 1\right)^{2}}
Ma_{0}^{2}\exp\left[2\sigma\tau + 2\mu(\eta-1)\right].
 \label{eq12}
\end{equation}
%=======================================================================%
Here we take into account that $\mu=\sqrt{\sigma Re}$. A complete
solution to Eq. (\ref{eq11}) is a superposition of the particular
solution Eq. (\ref{eq12}) and the solution to an auxiliary
equation
%=========================== Equation (13) ==============================%
\begin{equation}
\frac{\partial \vartheta}{\partial \tau} = \frac{1}{PrRe}
\frac{\partial^{2}\vartheta}{\partial \eta^{2}} ,
 \label{eq13}
\end{equation}
%=======================================================================%
which also grows in time as $\vartheta \propto \exp (2\sigma \tau )$
so that
%=========================== Equation (14) ==============================%
\begin{equation}
2\sigma Pr Re \vartheta = \frac{\partial^{2}\vartheta}{\partial
\eta^{2}} .
 \label{eq14}
\end{equation}
%=======================================================================%
Solution to the auxiliary Eq. (\ref{eq14}) may be found as
%=========================== Equation (15) ==============================%
\begin{equation}
\vartheta_{a} = \left[ A_{1} \exp (k\eta) + A_{2} \exp
(-k\eta)\right] \exp (2\sigma \tau ),
 \label{eq15}
\end{equation}
%=======================================================================%
with unknown amplitudes $A_{1}$, $A_{2}$ and
%=========================== Equation (16) ==============================%
\begin{equation}
k=\sqrt{2\sigma Pr Re}=\sqrt{2Pr}\mu.
 \label{eq16}
\end{equation}
%=======================================================================%
The complete solution to Eq. (\ref{eq11}) is a superposition of
$\vartheta_{p}$ and $\vartheta_{a}$, Eq. (\ref{eq12}) and Eq.
(\ref{eq15}), satisfying the adiabatic boundary condition at the
wall and the symmetry condition at the channel axis
%=========================== Equation (17) =============================%
\begin{equation}
\frac{\partial \vartheta}{\partial \eta}=0, \qquad \eta=0,1.
 \label{eq17}
\end{equation}
%=======================================================================%
Since $\vartheta_{p}$  is exponentially small at the axis
$\eta=0$, we find  $A_{1}=A_{2}$. Within the same exponential
accuracy, the decaying term $A_{2} \exp (-k\eta)$ is negligible
close to the wall. Thus we find at the wall $\eta = 1$
%=========================== Equation (18) ==============================%
\begin{equation}
kA_{1} \exp k = \left(\gamma - 1\right)\frac{Pr}{\left(2-Pr\right)}
\frac{\left(\Theta - 1\right)^{2}\mu^{3}}{\left(\mu - 1\right)^{2}}
Ma_{0}^{2}
 \label{eq18}
\end{equation}
%=======================================================================%
which determines the complete solution close to the wall as
%=========================== Equation (19) ==============================%
\begin{equation}
\vartheta = \frac{\left(\gamma-1\right)Pr\left(\Theta -
1\right)^{2}\mu^{2}}{2\left(2-Pr\right)\left(\mu - 1\right)^{2}}
Ma_{0}^{2}\exp (2\sigma \tau) \left[ \frac{2\mu}{k}
\exp\left[k(\eta-1)\right]-\exp \left[2\mu(\eta-1)\right] \right].
 \label{eq19}
\end{equation}
%=======================================================================%
The temperature increase at the walls $\eta = 1$ is
%Then, tking into account  $\mu = \sqrt{\sigma Re}$, we find
%temperature increase at the wall $\eta = 1$
%=========================== Equation (20) ==============================%
\begin{equation}
\vartheta = \frac{\left(\gamma-1\right)}{2}
\frac{\left(\Theta-1\right)^{2}\mu^{2}}{\left(\mu - 1\right)^{2}}
\frac{\sqrt{Pr}}{\sqrt{2}+\sqrt{Pr}} Ma_{0}^{2} \exp (2\sigma \tau).
 \label{eq20}
\end{equation}
%=======================================================================%
Finally, as an evaluation, we consider the limit of large Reynolds
number  $\sigma = \Theta^{2}/Re$,  $\mu = \Theta$,  which leads to
the temperature increase
%=========================== Equation (21) ==============================%
\begin{equation}
 \vartheta = \frac{\left(\gamma - 1\right)}{2}\Theta^{2}
 \frac{\sqrt{Pr}}{\sqrt{2}+\sqrt{Pr}} Ma_{0}^{2}
\exp (2\sigma \tau)= \frac{\left(\gamma - 1\right)}{2}\Theta^{2}
\frac{\sqrt{Pr}}{\sqrt{2}+\sqrt{Pr}} U_{w}^{2}/c_{s0}^{2}.
 \label{eq21}
\end{equation}
%=======================================================================%
It is interesting that viscous heating at the wall, Eq.
(\ref{eq21}), does not depend on the Reynolds number. Combination
$Ma_{w}=Ma_{0} \exp(\sigma\tau)=U_{w}/c_{s0}$ determines the
running value of the Mach number for an accelerating flame, which
is different from the initial value  $Ma_{0}= U_{f}/c_{s0}$. As an
example, we take $\Theta=8$, $Pr = 0.7$, $\gamma = 1.4$, and find
the temperature increase due to viscous stress $\vartheta \approx
4.8 Ma_{w}^{2} = 4.8 U_{w}^{2}/c_{s0}^{2}$.

We compare this result with the heating due to the compression
wave found in \cite{Bychkov-Akkerman-2006} as
%=========================== Equation (22) ==============================%
\begin{equation}
1+ \vartheta = \left( 1+\frac{\gamma -
1}{2}(\Theta-1)Ma_{w}\right)^{2}.
 \label{eq22}
\end{equation}
%=======================================================================%
Note that the designation  $\vartheta$ has a different meaning in
the present calculations and in \cite{Bychkov-Akkerman-2006}. In
the case of small Mach numbers we find
%=========================== Equation (23) ==============================%
\begin{equation}
\vartheta = (\gamma - 1)(\Theta-1)Ma_{w}= (\gamma
-1)(\Theta-1)\frac{U_{w}}{c_{s0}}.
 \label{eq23}
\end{equation}
%=======================================================================%
Using the same numbers as before we obtain a temperature increase
in a compression wave $\vartheta = 2.8 U_{w}/c_{s0}$. Thus, for
$\Theta = 8$, $Pr = 0.7$, $\gamma = 1.4$, we should expect a
comparable temperature increase due to the compression wave and
due to viscous stress at the Mach number about
$Ma_{w}=U_{w}/c_{s0}\approx 0.6$, see Eqs. (\ref{eq21}),
(\ref{eq23}). Besides, heating due to the compression wave is
mostly a 1D process equally strong at the walls and at the channel
axis. As both effects work together, we find preferential heating
at the walls. However, we have to remember that the theory of
flame acceleration \cite{Bychkov.et.al-2005} works quantitatively
well only for a noticeably subsonic flow, $Ma = 10^{-3} -
10^{-2}$. At such small Mach numbers heating due to viscous stress
is negligible, since it scales as  $\vartheta \propto Ma_{w}^{2}$,
while heating in a compression wave has another scaling $Ma
\propto Ma_{w}$. At Mach numbers above 0.05 the theory
\cite{Bychkov.et.al-2005} provides only order-of-magnitude
estimates.
 Direct numerical simulations
of the present paper show that, at Mach numbers above 0.05, flame
acceleration goes much slower than in the exponential regime Eq.
(\ref{eq2}). The critical value $Ma_{w}=U_{w}/c_{s0}\approx 0.6$
obtained above is also an order-of-magnitude estimate; it is
beyond the rigorous limits of the developed theory. Still, the
result Eq. (\ref{eq21}) may be useful even at relatively high
values of the Mach number. For comparison, temperature increase
due to viscous stress in the stationary boundary layer produced by
a uniform flow $U_{a}$ is \cite{Schlichting}
%=========================== Equation (24) ==============================%
\begin{equation}
\frac{\Delta T}{T_{a}}  = \frac{\gamma - 1}{2} \sqrt{Pr}
\frac{U_{a}^{2}}{c_{sa}^{2}} .
 \label{eq24}
\end{equation}
%=======================================================================%
Here $T_{a}$ and $c_{sa}$ are the temperature and the sound speed in
the uniform flow $U_{a}$ far away from the wall. In the present
problem flow velocity in the main stream (e.g. at the axis) plays
qualitatively the same role as  $U_{a}$. Rewriting Eq. (\ref{eq21})
in the same form, we find
%=========================== Equation (25) ==============================%
\begin{equation}
\frac{\Delta T}{T_{a}}  = \frac{\gamma - 1}{2}
\frac{\sqrt{Pr}}{\sqrt{2}+\sqrt{Pr}} \frac{U_{a}^{2}}{c_{sa}^{2}}
.
 \label{eq25}
\end{equation}
%=======================================================================%
Here we have taken into account that velocity at the axis ahead of
an accelerating flame is related to the total burning rate as
$U_{a}=\Theta U_{w}$, see Eq. (\ref{eq5}). The difference between
Eqs. (\ref{eq24}) and (\ref{eq25}) is then simply in the numerical
factor $\sqrt{2}+\sqrt{Pr}$, which is about $2.25$ for $Pr = 0.7$.
An exponentially accelerating flame produces about half the
temperature increase at the wall compared to that from the
stationary boundary layer. Thus one would expect that the
temperature increase at $Ma_{w}=0.05$ and above is somewhere
between predictions of Eqs. (\ref{eq24}) and (\ref{eq25}). In
order to investigate the role of viscous heating at relatively
large values of the Mach number we performed direct numerical
simulations. We stress once more that both formulas (\ref{eq24}),
(\ref{eq25}) predict viscous heating independent of the Reynolds
number. This conclusion is of great help in numerical simulations,
since it allows investigating the process even for rather low
values of the tube width.

%=======================================================================%
%        III. Basic equations of the direct numerical simulations       %
%=======================================================================%
\section{Basic equations of the direct numerical simulations}

We performed direct numerical simulations of the 2-D hydrodynamic
and combustion equations including transport processes (thermal
conduction, diffusion, viscosity) and chemical kinetics with an
Arrhenius reaction. The equations read
%=========================== Equation (26) =============================%
\begin{eqnarray}
\frac{\partial\rho}{\partial{t}}+
\frac{\partial}{\partial{x}_{i}}\left(\rho{u}_{i}\right)=0,
 \label{eq26}
\end{eqnarray}
%=======================================================================%
%=========================== Equation (27) =============================%
\begin{eqnarray}
\frac{\partial}{\partial{t}}\left(\rho{u}_{i}\right)+
\frac{\partial}{\partial{x_{j}}}
\left(\rho{u}_{i}u_{j}+\delta_{ij}p-\gamma_{ij}\right)=0,
 \label{eq27}
\end{eqnarray}
%=======================================================================%
%=========================== Equation (28) =============================%
\begin{eqnarray}
\frac{\partial}{\partial{t}}
\left(\rho\varepsilon+\frac{1}{2}\rho{u}_{i}u_{i}\right)+
\frac{\partial}{\partial{x}_{j}} \left(\rho{u}_{j}h+\frac{1}{2}
\rho{u}_{i}u_{i}u_{j}+q_{j}-u_{i}\gamma_{ij}\right)=0,
 \label{eq28}
\end{eqnarray}
%=======================================================================%
%=========================== Equation (29) =============================%
\begin{eqnarray}
\frac{\partial}{\partial{t}} \left(\rho{Y}\right)+
\frac{\partial}{\partial{x}_{i}} \left(\rho{u}_{i}Y-\frac{\mu}{Sc}
\frac{\partial{Y}}{\partial{x}_{i}}\right)=
-\frac{\rho{Y}}{\tau_R}\exp\left(-E/R_pT\right),
 \label{eq29}
\end{eqnarray}
%=======================================================================%
where $Y$ is the mass fraction of the fuel mixture,
$\varepsilon=QY+C_{V}T$ is the internal energy, $h=QY+C_{p}T$ is
the enthalpy, $Q$ is the energy release in the reaction, $C_{V}$,
$C_{p}$ are the heat capacities at constant volume and pressure.
It was assumed that the heat capacities do not depend on the
chemical reaction. The stress tensor $\gamma_{ij}$ and the energy
diffusion vector $q_{j}$ take the form
%=========================== Equation (30) =============================%
\begin{eqnarray}
\gamma_{ij}=\mu\left(\frac{\partial{u}_{i}}{\partial{x}_{j}}+
\frac{\partial{u}_{j}}{\partial{x}_{i}}-\frac{2}{3}
\frac{\partial{u}_{k}}{\partial{x}_{k}}\delta_{ij}\right),
 \label{eq30}
\end{eqnarray}
%=======================================================================%
%=========================== Equation (31) =============================%
\begin{eqnarray}
q_{j}=-\mu\left(\frac{C_{p}}{Pr}\frac{\partial{T}}
{\partial{x}_{j}}+\frac{Q}{Sc}\frac{\partial{Y}}
{\partial{x}_{j}}\right),
 \label{eq31}
\end{eqnarray}
%=======================================================================%
where $\mu\equiv\rho\nu$ is the dynamic viscosity, $Pr$ and $Sc$
are the Prandtl and Schmidt numbers, respectively. To avoid the
Zeldovich (thermal-diffusion) instability we take the unit Lewis
number  $Le \equiv Pr/Sc = 1$, with  $Pr = Sc = 0.75$. The
dynamical viscosity is $\mu=1.7\times10^{-5}Ns/m^{2}$. The
fuel-air mixture and burnt matter are assumed to be a perfect gas
with a constant molar mass  $m=2.9\times10^{-2}kg/mol$, with
$C_{V}=5R_{p}/2m$, $C_{p}=7R_{p}/2m$, where
$R_{p}\approx8.31J/(mol\cdot{K})$ is the perfect gas constant. The
adiabatic index is $\gamma \equiv C_{P}/C_{V}=1.4$. The equation
of state is
%=========================== Equation (32) =============================%
\begin{eqnarray}
  P=\rho{R}_{p}T/m.
  \label{eq32}
\end{eqnarray}
%=======================================================================%
We used the initial temperature of the fuel mixture $T_{f}=300 K$,
initial pressure $P_{f}=10^{5}Pa$, and initial expansion ratio
$\Theta_{0} = 8$. The initial Mach number is $Ma_{0}= U_{f}/c_{s0}
= 0.04$. Eq. (\ref{eq29}) describes a single irreversible reaction
of the first order, where the reaction rate obeys the Arrhenius
law with the activation energy $E_{a}$ and the factor of time
dimension $\tau_{R}$. In the simulations, the value of $U_{f}$ is
determined by the choice of thermal-chemical parameters such as
$E_{a}$, $\tau_{R}$, and the energy release in the reaction $Q =
(\Theta_{0}-1)C_{P}T_{f}$. In order to have better resolution of
the reaction zone, we took $E_{a}/R_{p}T_{f}=32$ in the
simulations. Note that the process of viscous heating addressed in
our paper does not depend on the activation energy. The factor
$\tau_{R}$ was adjusted to obtain a particular value of the planar
flame velocity $U_{f}$ by solving the associated eigen-value
problem. The flame thickness is conventionally defined as
%=========================== Equation (33) =============================%
\begin{eqnarray}
  L_{f}\equiv \frac{\mu}{Pr \rho_{f} U_{f}},
  \label{eq33}
\end{eqnarray}
%=======================================================================%
where $\rho_{f}= 1.16 kg/m^{3}$  is the initial mixture density.
However, we would like to stress that the value (\ref{eq33}) is
just a thermal-chemical parameter of length dimension in the
problem; the characteristic size of the burning zone may be an
order of magnitude larger \cite{Akkerman.et.al-2006b}.

In the studies of flame acceleration
\cite{Bychkov.et.al-2005,Akkerman.et.al-2005}, the main parameter of
simulations was the tube width  $2R$, which controlled the Reynolds
number. In the present paper we are interested in thermal heating
due to viscous stress. According to the present theory and the
classical results \cite{Schlichting}, viscous heating does not
depend on the Reynolds number (or this dependence is extremely
weak). For this reason, to study the effect, it is sufficient to
perform simulations only for one channel width. The channel width
was set to  $2R = 40 L_{f}$, which implied the Reynolds number
related to the planar flame velocity $Re = U_{f}R/\nu = R/L_{f}Pr =
26.7$. The Reynolds number related to the flow $Re = \left\langle
u_{z} \right\rangle 2R/\nu$ may be larger by several orders of
magnitude due to flame acceleration and thermal expansion of the
burning gas. We took non-slip and adiabatic boundary conditions at
the channel walls:
%=========================== Equation (34) =============================%
\begin{equation}
\textbf{u}=0, \qquad \hat{\textbf{n}}\cdot\nabla{T}=0.
 \label{eq34}
\end{equation}
%=======================================================================%
where $\hat{\textbf{n}}$ is a normal vector at the wall. At the
open channel end non-reflecting boundary conditions are used; the
boundary conditions were tested in \cite{Akkerman.et.al-2006b}. We
used two types of initial conditions: 1) a hemispherical flame
"ignited" at the channel axis at the closed end of the channel, or
2) a planar flame front.

The internal structure of the flame imitated the analytical
solution by Zel'dovich and Frank-Kamenetskii
\cite{Zeldovich.et.al-1985}. Particularly, for the hemispherical
case it was given by
%=========================== Equation (35) =============================%
\begin{equation}
T=T_{f} + T_{f}(\Theta-1)\exp\left( -
\sqrt{x^{2}+z^{2}}/L_{f}\right), \qquad \textrm{if} \qquad
z^{2}+x^{2}> r_{f}^{2}.
 \label{eq35}
\end{equation}
%=======================================================================%
%=========================== Equation (35a) =============================%
\begin{equation}
T=\Theta T_{f}, \qquad \textrm{if} \qquad z^{2}+x^{2}< r_{f}^{2}.
 \label{eq35a}
\end{equation}
%=======================================================================%
%=========================== Equation (35b) =============================%
\begin{equation}
Y =\frac{\Theta  - T/T_{f}}{\Theta -1}, \qquad P=P_{f}, \qquad
\textbf{u}=0.
 \label{eq35b}
\end{equation}
%=======================================================================%
Here  $r_{f}$ is the radius of initial flame ball at the closed
end of the channel. For an initially planar flame shape Eq.
(\ref{eq35}) is replaced by
%=========================== Equation (35c) =============================%
\begin{equation}
T=T_{f} +
T_{f}(\Theta-1)\exp\left[-\left(z-z_{0}\right)/L_{f}\right],
 \label{eq35c}
\end{equation}
%=======================================================================%
if $z > z_{0}$, with Eq. (\ref{eq35a}) for $z < z_{0}$ ($z_{0}$
denotes the initial flame position).

We simulated flame dynamics and DDT using a 2-D Cartesian
Navier-Stokes solver with chemical reactions developed at Volvo
Aero. The code is based on the cell-centered finite-volume scheme
developed by L.-E. Eriksson \cite{Eriksson-1993,Eriksson-1995}.
The numerical method has proved to be robust for modelling
different kinds of complex reacting flows. The TVD-limiting is
used where applicable to prevent overshoots in flow field
properties in regions of high gradients. The code has been
validated by solving various hydrodynamic problems
\cite{Eriksson-1993,Eriksson-1995,Eriksson-1987}, and was utilized
successfully in simulations of laminar flames at different
conditions of burning
\cite{Bychkov.et.al-2005,Akkerman.et.al-2005,Liberman.et.al-2006,
Akkerman.et.al-2006b,Bychkov&Liberman-2000,Bychkov.et.al-1996,
Travnikov.et.al-2000,Petchenko.et.al-2006,Petchenko.et.al-2007,
Liberman.et.al-2003}.

\begin{figure}
\subfigure[]{\includegraphics[width=\textwidth\centering]{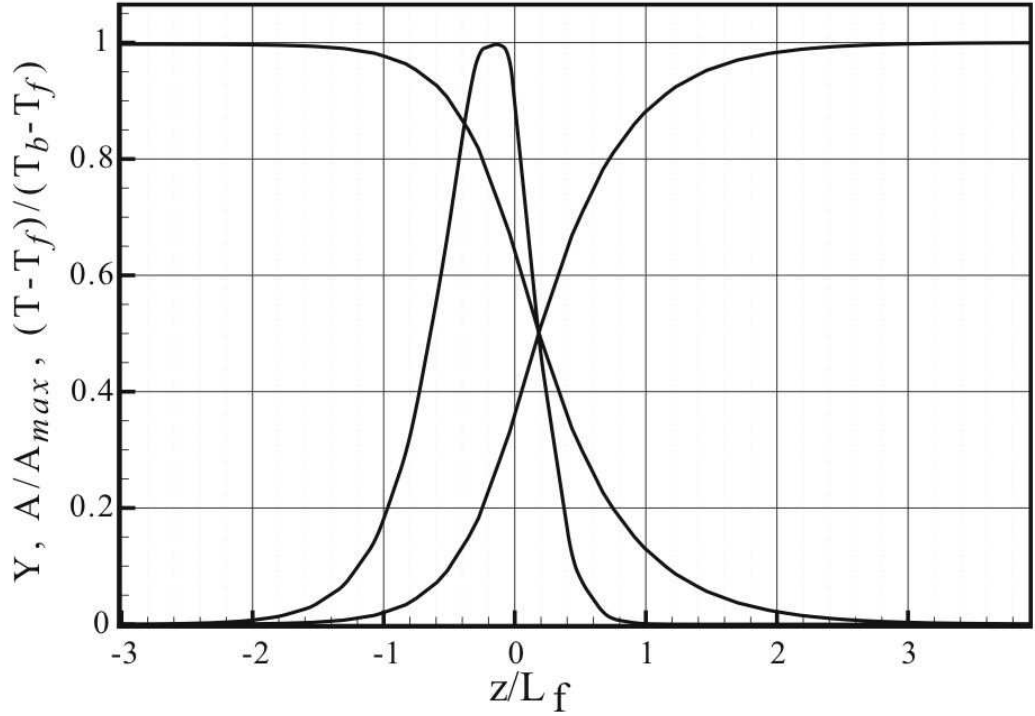}\label{fig-3a}}
\subfigure[]{\includegraphics[width=\textwidth\centering]{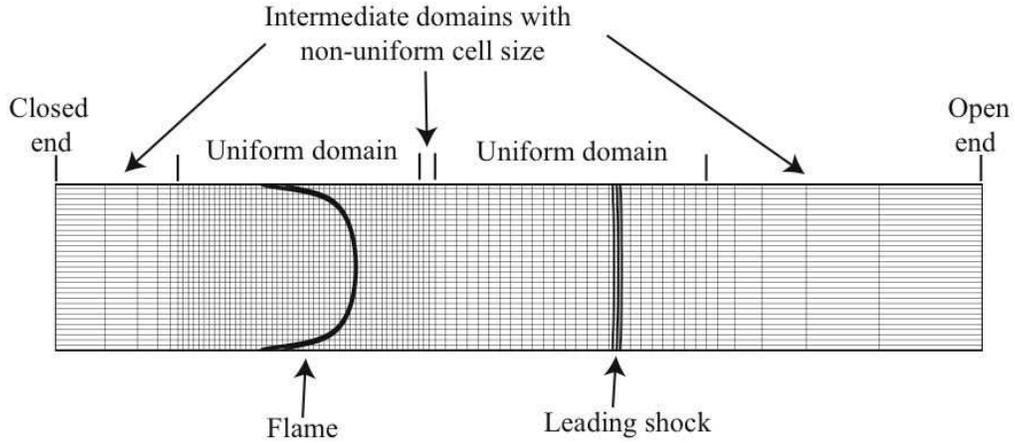}\label{fig-3b}}
 \caption{Illustration of the grid. (a) Profiles of the scaled
 temperature $(T-T_{f})/(T_b-T_{f})$, the mass fraction $Y$  and the reaction rate
 $A = (\rho Y / \tau _{R})\exp \left( - E_{a} / R_{p} T \right)$
 scaled by its maximal value within the flame front. (b) The adaptive
 grid used in numerical simulations.}
 \label{fig-3}
\end{figure}

The grids used in present simulations were non-uniform with higher
resolution around the flame front and pressure waves and lower
resolution in other parts of the calculation domain. Grid
resolution varied gradually from one zone to another along the
tube. Using this method we avoided processing a too large number
of cells. Square cells of the size $0.2L_{f} $ were used in the
flame domain to ensure isotropic propagation of the curved flame
in x and y directions. In \cite{Akkerman.et.al-2006b} we
demonstrated that such a grid is sufficient to resolve even a
strongly curved flame front. As an illustration, in Fig.
\ref{fig-3a} we present the profiles of the main variables inside
the flame front: the temperature distribution $(T-T_{f}) /
(T_{b}-T_{f}) $, the mass fraction $Y$ and the reaction rate $A =
({\rho Y}/\tau _{R})\exp \left( - E_{a} / R_{p} T \right)$ scaled
by its maximal value. We observe that the thermal--chemical
parameter $L_{f} $ defined by Eq. (\ref{eq33}) is much smaller
than the real flame thickness. According to Fig. \ref{fig-3a}, the
characteristic length scale of the temperature profile is about
$(4 - 5)L_{f} $, while the width of the active reaction zone is
about $L_{f} $. Taking the grid size of $0.2L_{f} $ in
$z$-direction we obtained about five grid points inside the active
reaction zone and 20--25 grid points inside the flame.

A sketch of the calculation grid is shown in Fig. \ref{fig-3b}.
The calculation grid is modified periodically in the course of
calculation, so that the sizes and positions of all domains in
Fig. \ref{fig-3b} change dynamically, following the flame front
and pressure waves. Thus the leading compression wave can never
reach the open tube end during a calculation. Cell size was
$0.4L_{f} $ around the front of the leading pressure wave. Splines
of the third order are used for re-interpolation of the flow
variables during periodic grid reconstruction to preserve the
second order accuracy of the numerical scheme. We performed
several test simulation runs with different resolution between 0.1
and 0.2$L_{f} $ in the flame domain. The tests demonstrated that
the difference in the flame velocity, pre-detonation time and
distance does not exceed 10\% for different resolutions.

%========================================================================%
%                      IV. Simulation results                            %
%========================================================================%
\section{Simulation results }

\begin{figure}
\subfigure[]{\includegraphics[width=\textwidth\centering]{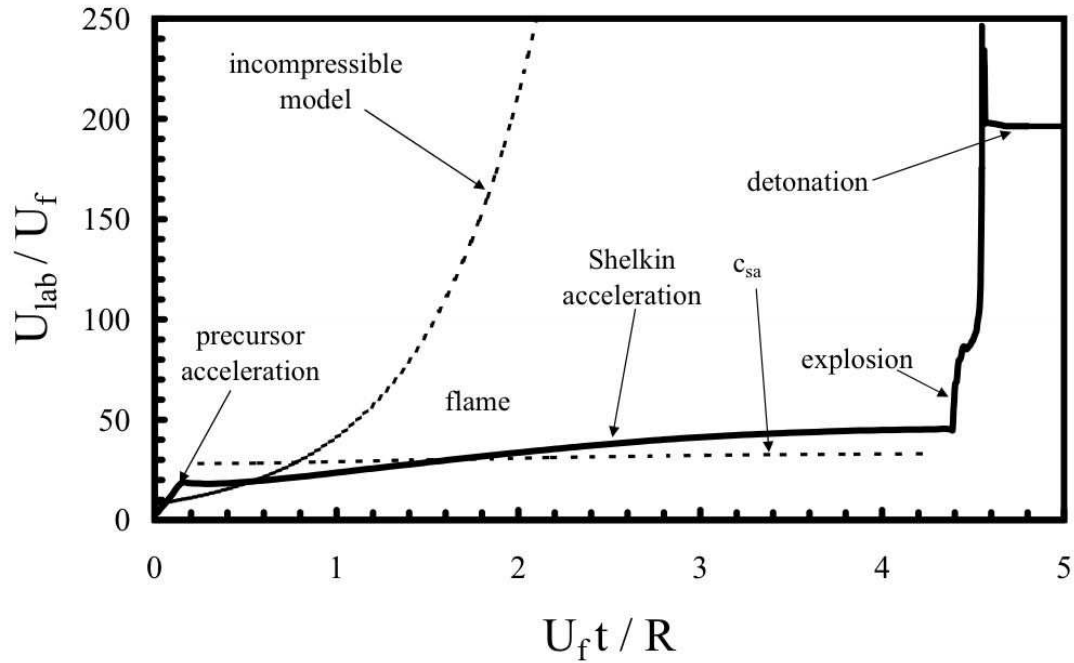}\label{fig-4a}}
\subfigure[]{\includegraphics[width=\textwidth\centering]{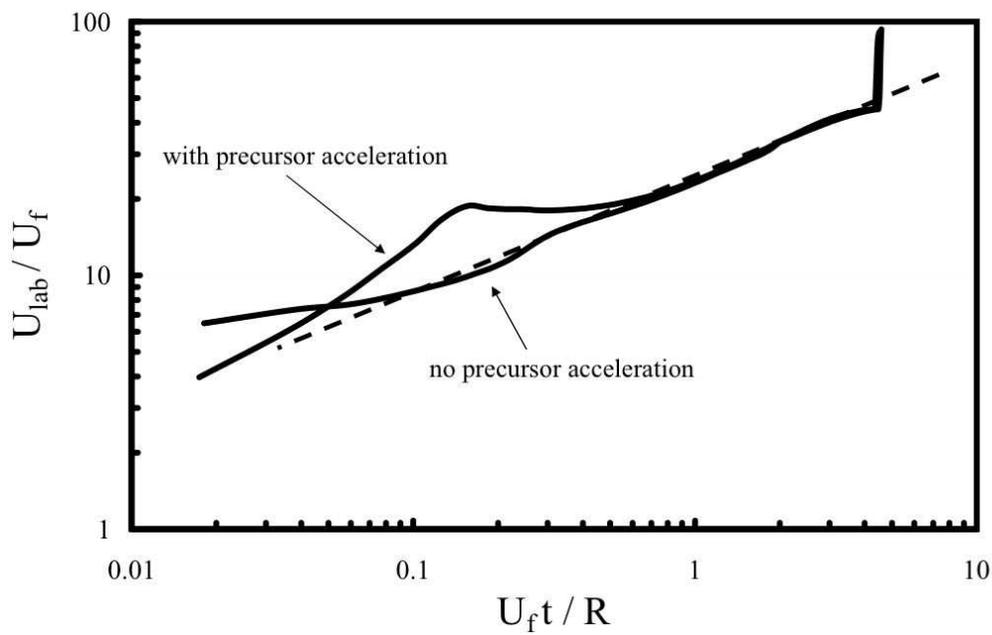}\label{fig-4b}}
 \caption{(a) Scaled velocity of the reaction front
(flame-explosion-detonation) in the laboratory reference frame,
$U_{lab}/U_{f}$. One dashed line presents theoretical predictions
for the incompressible flow. The other dashed line shows scaled
sound speed,  $c_{sa}/U_{f}$. (b) The same plot in log-log
variables for a flame ignited at the channel axis (with precursor
acceleration) and as a planar front (no precursor acceleration).}
 \label{fig-4}
\end{figure}

We performed direct numerical simulation of flame dynamics and
detonation triggering in a tube of width $2R = 40 L_{f}$  with
non-slip walls. The flame propagated from the closed tube end to
the open one; a semi-infinite tube was imitated by using
non-reflecting boundary conditions at the open end and by
adjusting the mesh to the leading pressure wave. We started the
simulations either with a hemispherical flame ignited at the tube
axis or with a planar flame front. Fig. \ref{fig-4} shows velocity
of the combustion front in the laboratory reference frame of the
tube, $U_{lab}/U_{f}$, in the case of hemispherical ignition; the
velocity was calculated as the average over the reaction front.
Flame dynamics involves several velocity parameters,
which have quite different physical meaning and may differ
quantitatively by one or two orders of magnitude. First, we have a
planar flame velocity $U_{f}$, which characterizes local normal
velocity of the flame front relative to the moving fuel mixture
(at least in the model of an infinitely thin front). The value $U_{f}$
works as a standard velocity scaling in the
problems of flame dynamics. To be accurate, $U_{f}$ depends on
thermal and chemical parameters of the fuel mixture, which vary in
the process of detonation triggering. Here $U_{f}$ designates the
initial planar flame velocity. Second, we have the burning rate
$U_{w}$, which specifies total amount of fuel mixture burnt per
unit time (see also Section II). The burning rate $U_{w}$ may
exceed $U_{f}$ by several orders of magnitude in the process of
flame acceleration. Still, the value $U_{w}$ does not describe
flame propagation in any particular reference frame except for the
case of stationary or statistically-stationary curved flames like
in
\cite{Bychkov.et.al-1996,Travnikov.et.al-2000,Petchenko.et.al-2006b,
Liberman.et.al-2003,Kadowaki-Hasegawa-2005}.

Flame propagating from the closed end
pushes a strong flow in the fuel mixture. In the incompressible
case, average velocity of the flow is $\left\langle
u_{z}\right\rangle = (\Theta-1)U_{w}$, where averaging is
performed over the channel cross-section. The flow drifts the
flame, and the drift velocity is different for different parts of
the flame front. For this reason, the burning rate $U_{w}$ is also
different from flame velocity in the laboratory reference frame,
with respect to the tube walls,  $U_{lab}$. We stress that Fig.
\ref{fig-4} presents flame velocity in the laboratory reference
frame, $U_{lab}$, not the burning rate  $U_{w}$. In the case of an
essentially subsonic flame, we have a simple relation
$U_{lab}=\Theta U_{w}$, which may also provide an estimate in the
case of finite Mach numbers. Different velocity parameters imply
different Mach numbers as well, which we designate by
$Ma_{0}=U_{f}/c_{s0}$, $Ma_{w}=U_{w}/c_{sa}$,
$Ma_{lab}=U_{lab}/c_{sa}$; label "$a$" refers to the position at
the channel axis just ahead of the flame front. Taking a rather
low initial Mach number related to the planar flame velocity
$Ma_{0}=0.04$ with  $\Theta=8$, we have immediately
$Ma_{lab}=0.32$, which is not so low. Soon after ignition, in a
short time about $(0.2-0.3)R/U_{f}$, burning rate increases by
order of magnitude in the process of precursor acceleration
\cite{Clanet-Searby-1996,Bychkov.et.al-2007}. As a result, quite
quickly the Mach number $Ma_{lab}$ becomes comparable to unity. We
also point out that  $Ma_{w}$, $Ma_{lab}$ take into account
variations of the sound speed in the fuel mixture in time and
space; scaled sound speed $c_{sa}/U_{f}$ is shown by the dashed
line in Fig. \ref{fig-4a}.

Transition from one combustion regime to another
(flame-explosion-detonation) may be clearly seen by drastic
changes in the front velocity. In the flame regime, we can see
front acceleration, which involves two different physical
mechanisms. First, we observe a short period of precursor
acceleration, related to the hemispherical geometry of flame
ignition. This type of acceleration is possible even in the case
of friction-free walls. Precursor acceleration goes noticeably
faster than the Shelkin mechanism, but this process is limited in
time. Precursor acceleration for a cylindrical tube was
investigated experimentally in \cite{Clanet-Searby-1996} and
numerically in \cite{Bychkov.et.al-2007}. The paper
\cite{Bychkov.et.al-2007} presented also the analytical theory of
the process. In the present geometry of a 2-D channel the
precursor acceleration is somewhat different from the cylindrical
case; still, the basic features of the process remain the same.

The precursor acceleration is followed by acceleration due to the
Shelkin mechanism produced by thermal expansion and non-slip at
the walls. The theory of the Shelkin acceleration in
incompressible flows has been developed in
\cite{Bychkov.et.al-2005,Akkerman.et.al-2005}. However, in the
present case Mach number $Ma_{lab}$ is comparable to unity or
above, and the incompressible theory of flame acceleration
describes the process only qualitatively. A dashed line in Fig.
\ref{fig-4a} shows the exponential growth predicted by the
incompressible theory Eq. (\ref{eq2}); it goes well above the
numerical results as soon as the precursor acceleration is over.
Investigation of the compressibility effects on flame acceleration
is beyond the scope of the present paper; this is work for the
future. Still, as a preliminary result, Fig. \ref{fig-4b} shows
the flame velocity in log-log variables. In order to elucidate the
transition effects related to a particular geometry of flame
ignition, we have performed two simulation runs: that of a flame
ignited at the tube axis and one for a flame ignited as a planar
front. In the first case we have precursor acceleration of the
flame front, while in the second case this effect is missing. As
we can see, both simulation runs tend to the same asymptotic
regime of flame acceleration with $U_{lab}\propto(U_{f}t/R)^{n}$,
where $n\approx 0.36$. More detailed investigation of the process
will be presented elsewhere. It is well known that flame is
subsonic; still, this rule applies to the burning rate $U_{w}$,
not to $U_{lab}$. In Fig. \ref{fig-4b}, $U_{lab}$ crosses the
sonic line at the time instant about  $U_{f}t/R=30$. This time
instant is often called run-up time to supersonic flames
\cite{Kuznetsov.et.al-2005}.

\begin{figure}
\includegraphics[width=\textwidth\centering]{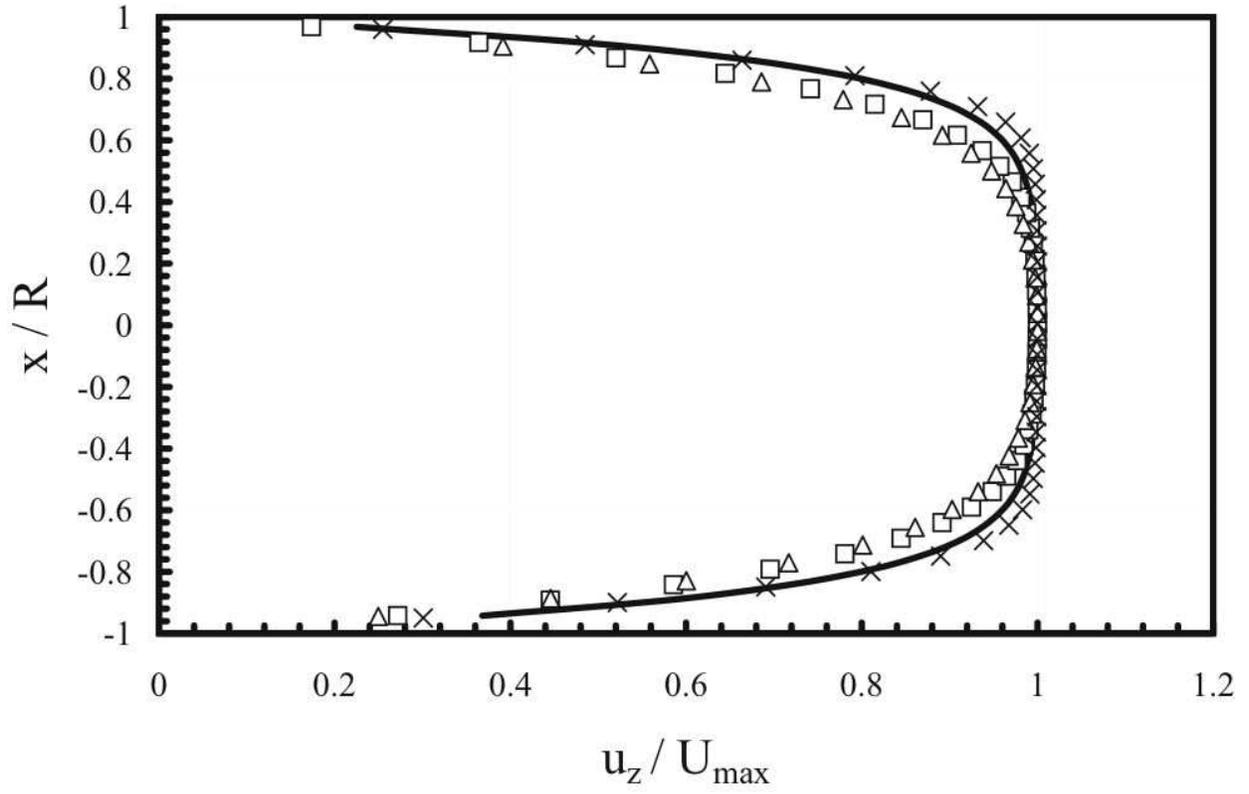}
 \caption{The velocity profile $u_{z}$  scaled by the amplitude
 $U_{max}$. The solid line shows the theoretical result Eq. (\ref{eq5}).
 The markers correspond to the present simulation results just ahead of the
flame front at the time instants $U_{f}t/R = 0.92; 1.74; 3.43$.}
 \label{fig-5}
\end{figure}

An accelerating flame produces a flow in the fuel mixture, which
is non-uniform because of the non-slip boundary conditions. A
characteristic velocity profile is shown in Fig. \ref{fig-5}: the
results of numerical simulations are compared to the predictions
of the incompressible theory \cite{Bychkov.et.al-2005}. The
velocity profile is calculated at different time instants just
ahead of the flame front and scaled by the maximum velocity value
at the tube axis. Both the theory and the present simulations
demonstrate the same qualitative structure of the flow, which
consists of an almost uniform stream close to the axis and
transitional (boundary) layers close to the walls. The velocity
profile  is in a good  agreement with the incompressible case, not
only qualitative, but even quantitative.
 A non-uniform viscous flow implies stress,
which is stronger close to the channel walls. As a consequence,
one should expect additional heating of the fuel mixture due to
the viscous stress, as discussed in Section II. In this Section,
our main goal is to investigate the process of viscous heating
numerically and to compare it to the theory of Section II.

\begin{figure}
\includegraphics[width=\textwidth\centering]{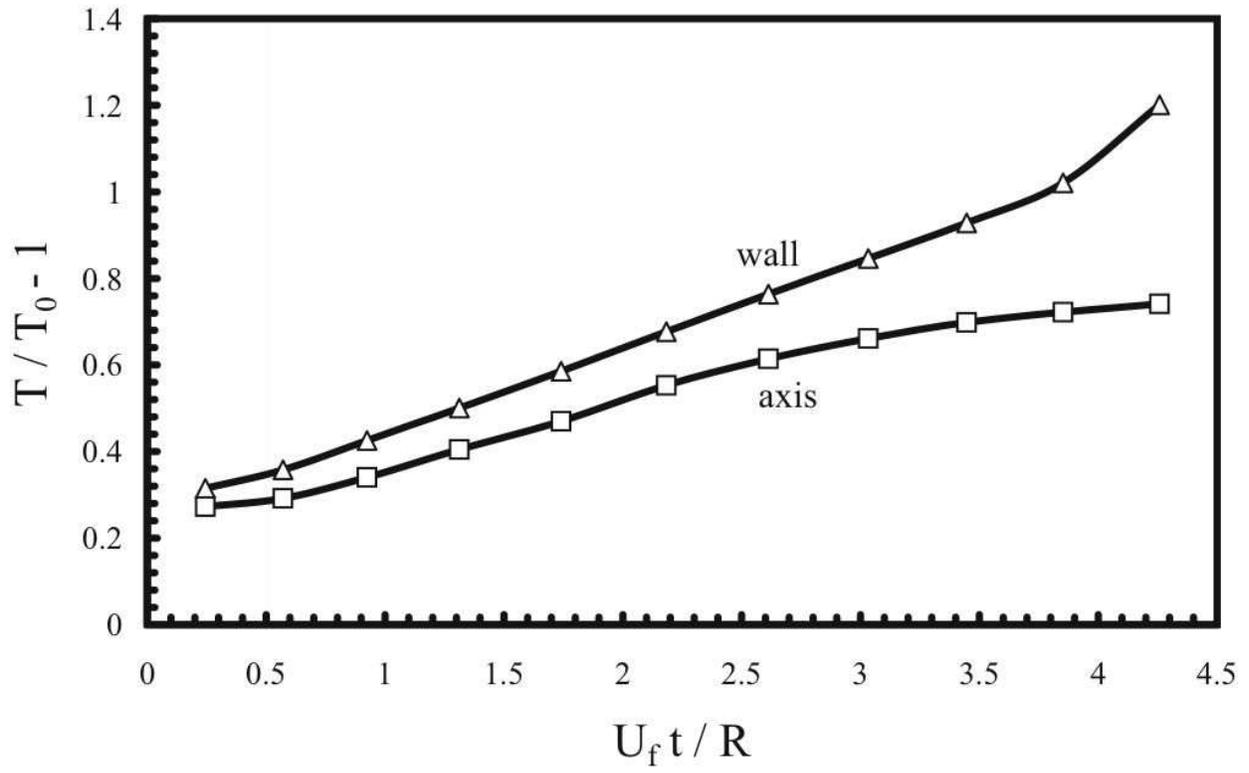}
 \caption{Temperature increase at the wall and at the channel axis versus time.}
 \label{fig-6}
\end{figure}

\begin{figure}
\includegraphics[width=\textwidth\centering]{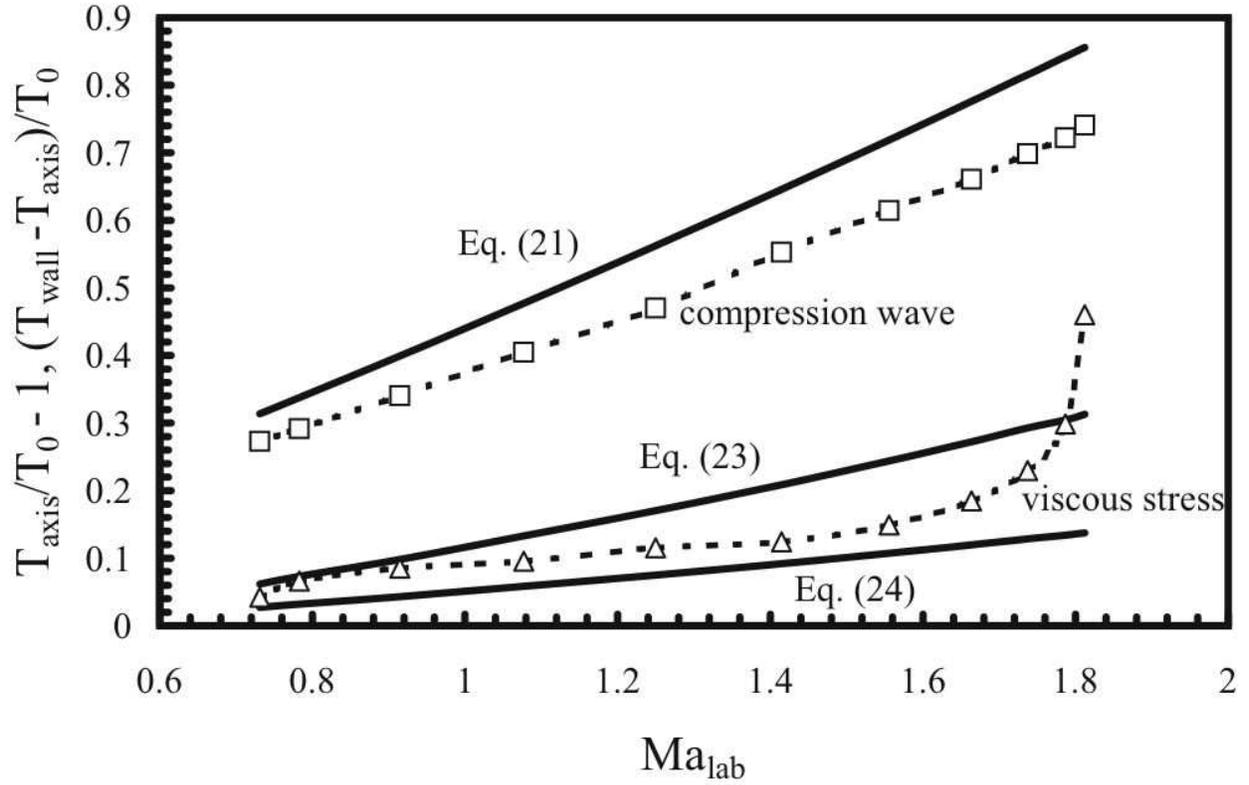}
 \caption{Temperature increase versus Mach number in the laboratory
reference frame. Solid lines show theoretical predictions for
temperature increase in the compression wave at the axis, Eq.
(\ref{eq22}), and additional increase due to viscous stress at the
wall, Eqs. (\ref{eq24}), (\ref{eq25}). The dashed lines and the
markers show respective numerical results. }
 \label{fig-7}
\end{figure}

\begin{figure}
\includegraphics[width=\textwidth\centering]{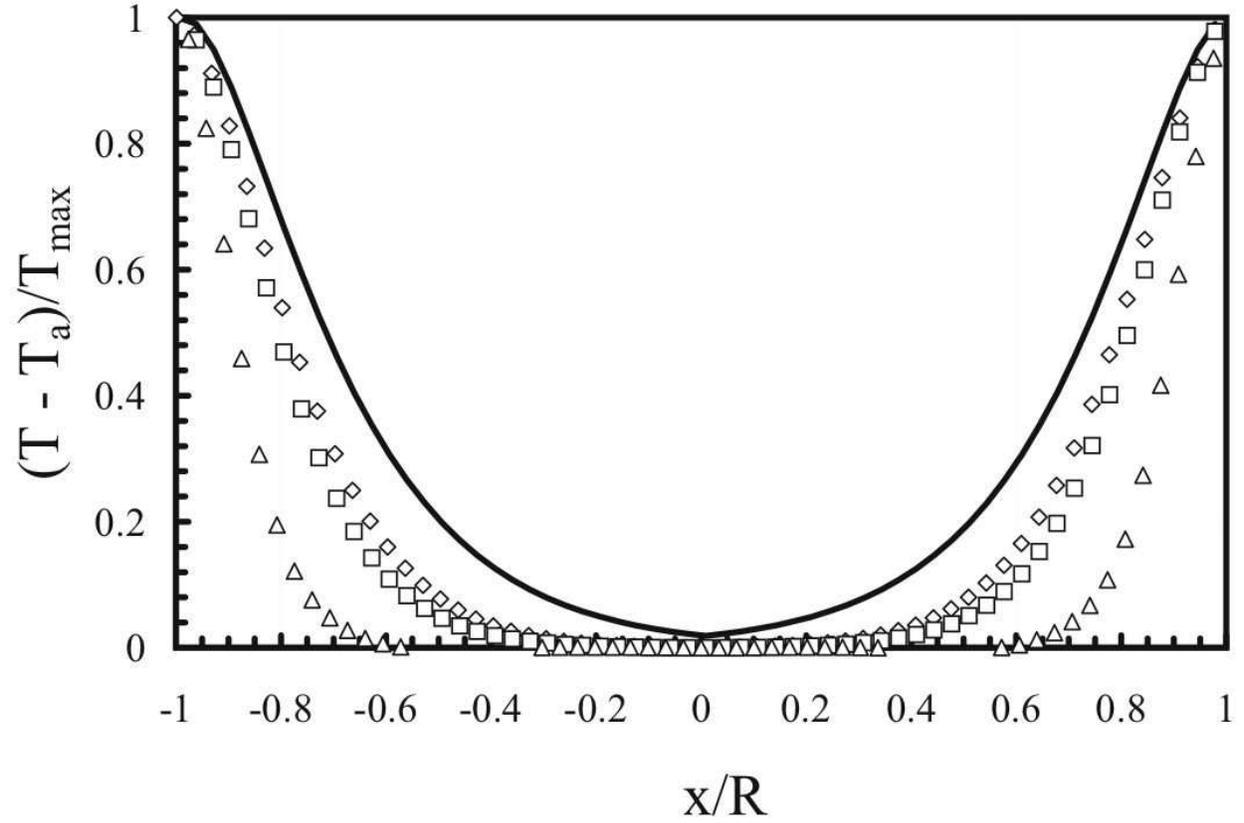}
 \caption{Scaled temperature profile  $(T-T_{axis})/(T_{wall}-T_{axis})$.
 The solid line shows the theoretical result Eq. (\ref{eq19}). The markers
 correspond to the present simulation results at the distances $(6; 425; 930) L_f$
 ahead of the flame front at the time instant $U_{f}t/R = 3.43$.}
 \label{fig-8}
\end{figure}

The temperature increase in the fuel mixture just ahead of the
flame front is shown in Fig. \ref{fig-6} as a function of time and
at two positions: at the tube axis and at the channel wall. We can
see that, within a short time after ignition  $U_{f}t/R=0.2-0.3$,
the temperature increase is already different from zero, being
about $\Delta T/T_{0}\approx 0.25$. This happens because of the
leading shock wave, pushed by the flame in the process of
precursor acceleration. The flow velocity behind the shock is
about  $\Delta u_{z}/U_{f}\approx \Theta - 1$, and thermal
expansion is typically quite strong. Therefore, in spite of a
small initial Mach number $Ma_{0}=0.04$, the flow velocity is not
so small in comparison with sound speed even for a planar flame
front,  $\Delta u_{z}/c_{s0}=0.28$. Precursor acceleration makes
this value even larger. At the beginning, temperature increase at
the walls and at the tube axis is practically the same, which
illustrates almost 1D structure of the compression wave. As time
goes on, the flame accelerates, the Mach number of the flow
increases and heating due to viscous stress becomes noticeable in
agreement with the theory of Section II. By the time of explosion,
$U_{f}t/R \approx 4.38$, difference between temperature at the
axis and at the walls is noticeable; still, this difference is not
as strong as predicted by the theory. In order to compare the
numerical simulations and the theory quantitatively, in Fig.
\ref{fig-7} we plot temperature increase versus Mach number
$Ma_{lab}$. In that case we separate heating in the compression
wave (at the axis), and the effect of viscous stress as
temperature difference between the wall and the axis,  $T_{wall}-
T_{axis}$. Heating in the compression wave at the axis is shown by
squares. Temperature at the axis increases almost linearly with
the Mach number, in agreement with the theory, Eq. (\ref{eq22}),
presented by the solid line. The theoretical prediction holds for
a 1D isentropic flow; small deviations of the theory and
simulations indicate that the flow is not completely 1D and
isentropic. Two other solid lines show theoretical predictions for
temperature increase due to viscous stress ahead of an
exponentially accelerating flame, Eq. (\ref{eq25}), and in a
stationary boundary layer, Eq. (\ref{eq24}). These two formulas
present two limiting cases of strong acceleration and no
acceleration at all. As we pointed out, at relatively high values
of the Mach number, the flame acceleration is much slower than
exponential. By this reason it is natural to expect that the
numerical points should be somewhere in between the predictions of
Eqs. (\ref{eq24}) and (\ref{eq25}). Indeed, Fig. \ref{fig-7} shows
that this is the case; the numerical results come rather close to
that of Eq.~(\ref{eq25}), being a little above the theoretical
curve. Only at the end of the simulation run, temperature at the
wall increases drastically, which indicates the beginning of the
explosion. It is also interesting to look at the temperature
distribution in the channel. Fig.~\ref{fig-8} compares temperature
profiles at different distances ahead of the flame front at the
time instant $U_{f}t/R = 3.43$ to the theoretical result
Eq.~(\ref{eq19}). The theoretical predictions look quite similar
to the numerical results: all the time we observe the effect of
viscous heating well-localized at the walls. Still, the theory
predicts a self-similar temperature profile independent of
position, since it was developed for an exponentially accelerating
flame in an incompressible flow. In the numerical simulations
flame acceleration is not exponential, the flow is not
incompressible, and the temperature profile changes in space and
time. As we can see in Fig.~\ref{fig-8}, the transitional zone of
viscous heating is wider close to the flame front, and it becomes
more localized far away from the flame.

\begin{figure}
\includegraphics[width=\textwidth\centering]{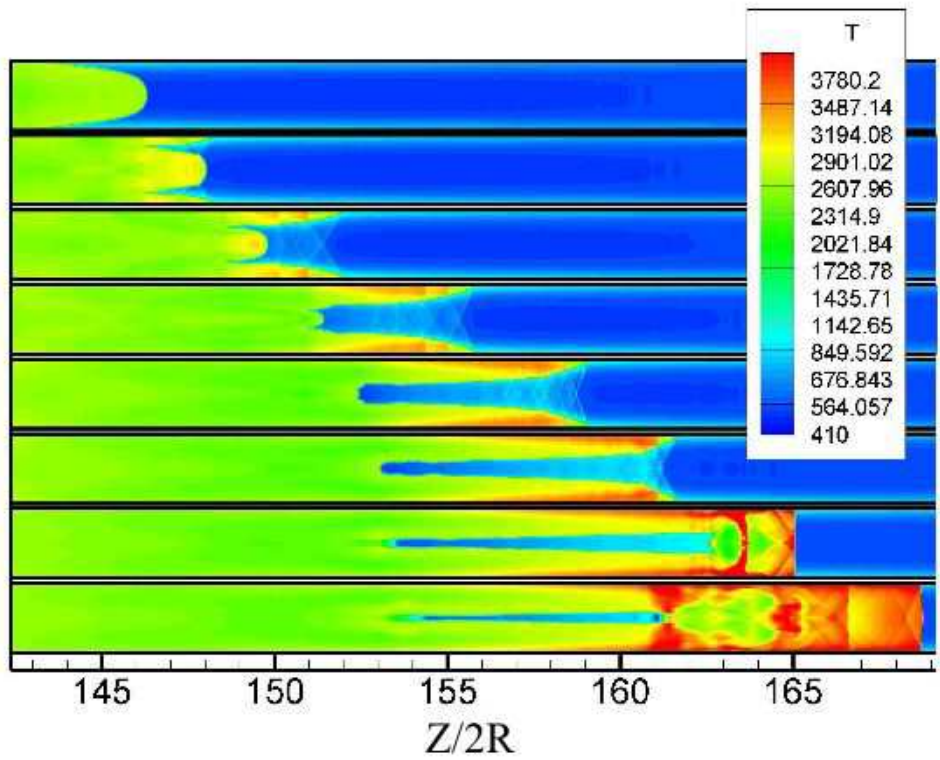}
 \caption{Temperature field during DDT for the time period from $U_{f}t/R = 4.38$
 to $U_{f}t/R = 4.58$ with equal time intervals between the snapshots.}
 \label{fig-9}
\end{figure}

As we can see, heating due to viscous stress is not large in
comparison with that in a compression wave, and, in principle,
explosion theoretically could be possible even without viscous
heating. Fig.~\ref{fig-9} presents temperature field snapshots in
the area of the accelerating flame from beginning of the explosion
till detonation. Prior to the explosion, we can see light blue
strips along the walls indicating regions of viscous heating.
Still, Arrhenius reactions are extremely sensitive to temperature;
even this slight non-uniformity  influences strongly the structure of the
explosion.  The explosion may be observed already at the second
snapshot: it develops in the form of two narrow tongues spreading
fast along the tube walls from the flame front to the fuel
mixture. Temperature in the explosion zone is noticeably higher
than it was behind the flame front. The tongues grow in length and
width, until the original flame front becomes engulfed in the
explosion. However, even at that time the explosion front remains
essentially multi-dimensional: a large amount of the unburned
matter separates two explosion tongues. The explosion pushes
inclined pressure waves, well seen in Fig.~\ref{fig-9} by the
light blue color. Intersection of the pressure waves increases
temperature at the channel axis, reduces the reaction time, until
a bridge is formed between two tongues of the explosion. As two
tongues meet, they soon develop into detonation, which can be seen
in the last two snapshots of Fig.~\ref{fig-9}. Meeting of two
explosion tongues may be also seen in Fig.~\ref{fig-4}. as an
extremely sharp peak in the velocity dependence. Strictly
speaking, this peak is infinitely sharp and high, since it
corresponds to a singularity of pocket formation. In
Fig.~\ref{fig-4}, we cut down the peak to make the plot more
illustrative. In the present calculations, DDT distance is $330
R$, but this value is, of course, sensitive to a particular
chemical kinetics.

%========================================================================%
%                     V. Summary and Discussion                          %
%========================================================================%
\section{Summary}

The purpose of the present paper was to investigate the role of
viscous heating in DDT. We solved this problem both analytically
and by direct numerical simulations. The analytical theory was
developed within the approximation of incompressible flow, which
holds at sufficiently low values of the Mach number. On the
contrary, the numerical simulations were performed for relatively
large Mach numbers. Both theory and numerical simulations
demonstrated that the role of viscous heating increases with the
Mach number. Viscous stress makes temperature higher at the
channel walls in comparison with the axis. Temperature profile in
the channel is given by Eq. (\ref{eq19}), temperature increase at
the wall due to viscous stress is determined by Eq. (\ref{eq25}).
The theoretical results agree rather well with the numerical
simulations, though, when making a comparison, one has to take
into account much slower flame acceleration at larger values of
the Mach number. Even just before the explosion, additional
temperature increase due to viscous heating is noticeably smaller
than the temperature increase in the compression wave. Thus,
viscous heating should be treated only as a supplementary process
in DDT. In principle, DDT could be possible with heating provided
by the compression wave only in line with the Shelkin scenario
\cite{Shelkin-1940,Roy.et.al-2004,Bychkov-Akkerman-2006}. Still,
neglecting viscous heating in a theoretical model like
\cite{Bychkov-Akkerman-2006} may lead to considerable disagreement
in the time and distance of detonation triggering predicted by the
theory and obtained in experiments/numerical simulations. Viscous
heating makes explosion triggering easier and faster because of
extreme sensitivity of Arrhenius reactions to temperature
variations. In a certain sense, it plays the same role as hot
spots: it is not necessary, but helpful. Non-uniform temperature
distribution leads to an essentially multi-dimensional structure
of the explosion, which pushes inclined pressure waves and
eventually develops into detonation.

%========================================================================%
%                          Acknowledgments                               %
%========================================================================%
\section{Acknowledgements}

This work was supported by the Swedish Research Council (VR).
Numerical simulations were performed at High Performance Computer
Center North (HPC2N), Ume\aa, Sweden, within SNAC project
007-07-25. The authors also wish to thank Arkady Petchenko and
Michael Modestov for useful discussions.

%========================================================================%
%                              References                                %
%========================================================================%

%===================================================================%
%                              Figures                              %
%===================================================================%
\newpage

%\centerline{FIGURE CAPTIONS}

%============================= Figure 1 ============================%
%============================= Figure 2 ============================%
%============================= Figure 3 ============================%

%============================= Figure 4 ============================%
%============================= Figure 5 ============================%
%============================= Figure 6 ============================%
%============================= Figure 7 ============================%
%============================= Figure 8 ============================%
%============================== Figure 9 ===========================%
%===================================================================%

\end{document}